\documentclass[10pt,conference,letter]{IEEEtran}
\IEEEoverridecommandlockouts
\usepackage{fancyhdr}
\usepackage{comment}
\usepackage{algorithm2e}
\usepackage{amsmath}
\usepackage{pifont}
\usepackage{cite}
\ifCLASSINFOpdf
  \usepackage[pdftex]{graphicx}
  \graphicspath{{./figures/}{./graphs/}}
  \DeclareGraphicsExtensions{.pdf,.jpeg,.png}
\else
\fi
\ifCLASSOPTIONcompsoc
  \usepackage[caption=false,font=normalsize,labelfont=sf,textfont=sf]{subfig}
\else
  \usepackage[caption=false,font=footnotesize]{subfig}
\fi
\usepackage{url}
\usepackage{multirow}
\newcommand{\MySection}[1]{
  \vspace{-0.3ex}
  \section{#1}
  \vspace{-0.3ex}
}
\newcommand{\MySubsection}[1]{
  \vspace{-0.5ex}
  \subsection{#1}
  \vspace{-0.3ex}
}
\newcommand{\MySubsubsection}[1]{
  \subsubsection{#1}
}

\usepackage{pifont}

\begin{document}
\title{\fontsize{20}{20}\selectfont
   Scalable Distributed DNN Training using TensorFlow and CUDA-Aware MPI:
Characterization, Designs, and Performance Evaluation
}
\author{
\IEEEauthorblockN{
       Ammar Ahmad Awan,
       Ching-Hsiang Chu,\\
       Hari Subramoni,
       and
       Dhabaleswar K. Panda
   }
   \IEEEauthorblockA{
           Department of Computer Science and Engineering \\
           The Ohio State University \\
           \{awan.10, chu.368, subramoni.1, panda.2\}@osu.edu 
   }
   
\and
\IEEEauthorblockN{
    Jeroen B\'edorf
}
\IEEEauthorblockA{
           Minds.ai \\
           Santa Cruz, the United States \\
           jeroen@minds.ai
   }
\IEEEauthorblockA{
           Leiden Observatory, Leiden University \\
           Leiden, the Netherlands
   }
}
\maketitle
\thispagestyle{fancy}
\pagenumbering{arabic}
\begin{abstract}
The current wave of advances in Machine Learning (ML) and Deep Learning (DL)
have been triggered by the availability of large-scale datasets, efficient CPU
and GPU hardware, and development of easy-to-use software frameworks like
TensorFlow (TF), Caffe and Torch. %
TensorFlow has been, by far, the most widely adopted ML/DL framework. However, 
little exists in the
literature that provides a thorough understanding of the capabilities which
TensorFlow offers for the distributed training of large ML/DL models that need
computation and communication at scale. Most commonly used distributed training
approaches for TF can be categorized as follows: 1) Google Remote
Procedure Call (gRPC), 2) gRPC+`X': X = (InfiniBand Verbs, Message Passing Interface (MPI), 
and GPUDirect RDMA), and 3) No-gRPC: Baidu Allreduce with MPI, Horovod with MPI, and
Horovod with NVIDIA NCCL.  
In this paper, we provide an in-depth performance characterization and analysis
of these distributed training approaches on various GPU clusters including
the Piz Daint system (\#6 on Top500). We perform experiments to gain
novel insights along the following vectors: 1) Application-level scalability of
DNN training, %
2) Effect of Batch Size on scaling
efficiency, 3) Impact of the MPI library used for no-gRPC approaches,
and 4) Type
and size of DNN architectures (e.g ResNet vs. MobileNet). Based on these
experiments, we present two key insights: 1) Overall, No-gRPC designs achieve
better performance compared to gRPC-based approaches for most configurations,
and 2) The performance of No-gRPC is heavily influenced by %
the gradient aggregation using the Allreduce communication pattern. Finally, we
propose a truly CUDA-Aware MPI\_Allreduce design that exploits 1) CUDA
kernels to perform large reductions on the GPU and 2) A pointer cache to avoid
overheads involved in queries to the CUDA driver. Our proposed designs have been
implemented in MVAPICH2-GDR and offer 5-17$\times$ better performance than NCCL2
for small and medium messages, and reduces latency by 29\% for large messages on
16 GPUs (nodes). The proposed optimizations help Horovod-MPI to achieve
approximately 90\% scaling efficiency for ResNet-50 training on 64 GPUs.
Further, Horovod-MPI achieves 1.8$\times$ and 3.2$\times$ higher throughput
than the native gRPC method for ResNet-50 and MobileNet, respectively, on the
Piz Daint cluster.
 \end{abstract}
\begin{figure}[htbp]
    \vspace*{-0.5\baselineskip}
    \includegraphics[width=0.9\columnwidth]{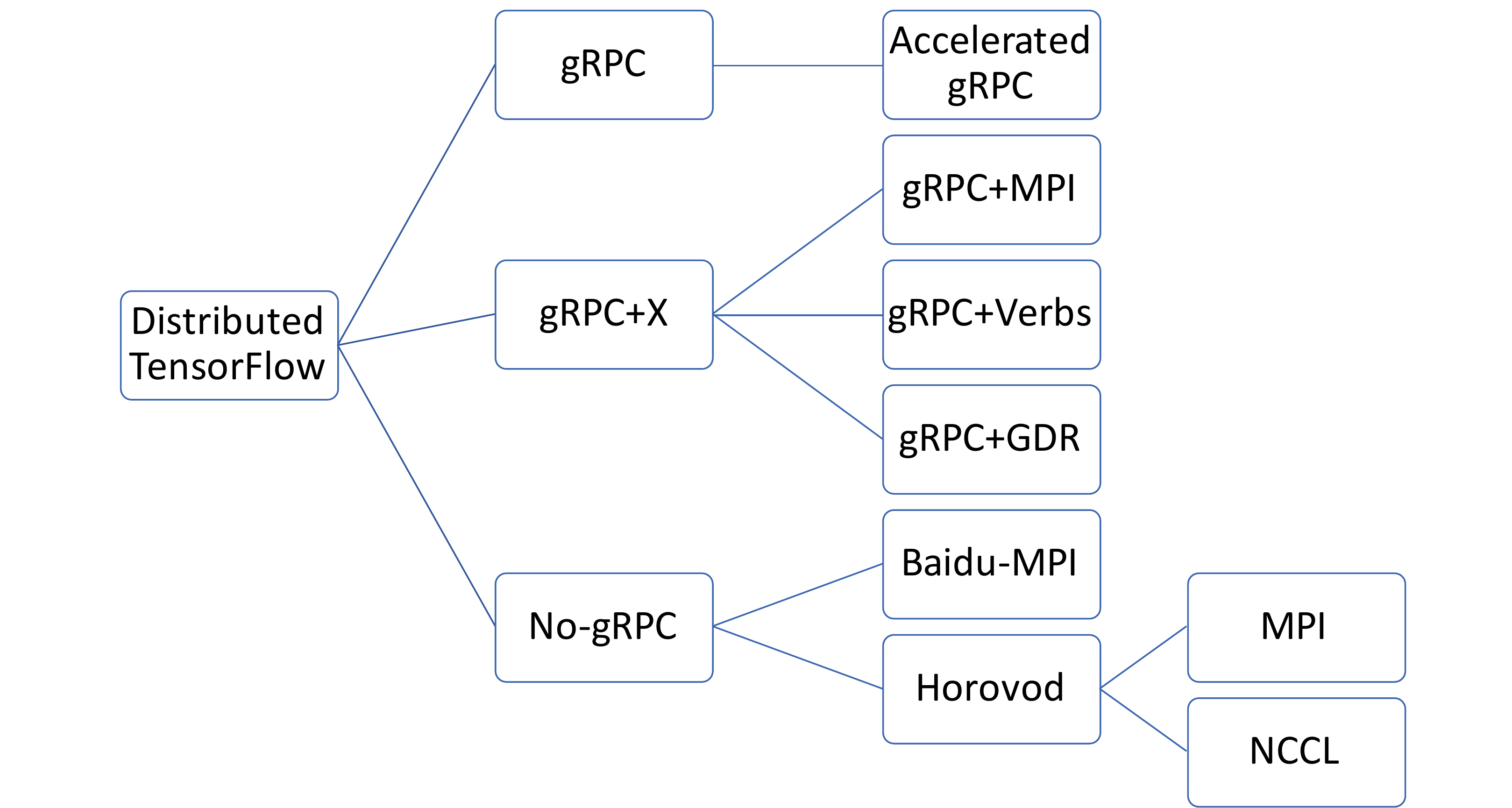}
    \vspace*{-0.5\baselineskip}
    \caption{Various Approaches for Distributed Training using TensorFlow}
    \vspace*{-0.5\baselineskip}
    \label{fig:overview}
\end{figure}

\MySection{Introduction}
\label{sec:intro}
Deep Learning (DL) has been a significant contributor to the recent achievements
in the Artificial Intelligence (AI) realm. Novel approaches like back-propagation
in Deep Neural Networks (DNNs) were investigated around the 1980 time frame~\cite{Rumelhart:1986:LIR:backpropagation}.
However, the potential of these approaches was marred by slow hardware and lack
of sufficient training data. To this end, the current resurgence and a renewed
interest in DL and DNN-based solutions to classical as well as new Machine
Learning (ML) problems can be attributed to the widespread availability of 1)
versatile and large-scale datasets like %
ImageNet~\cite{imagenet}, and 2) efficient
computation capabilities in modern hardware architectures like Graphics Processing Units (GPUs) and
multi-/many-core CPUs. These two trends have positively influenced the
development of several high-level DL toolkits like
Caffe~\cite{caffe,caffe2}, Microsoft Cognitive Toolkit (CNTK),
Facebook PyTorch~\cite{pytorch}, and Google TensorFlow~\cite{tensorflow:OSDI16}.
Implementing DNN and back-propagation techniques
in an efficient manner has been a challenging problem. However, these toolkits
have played a crucial role in making ML and DL more accessible for both
academic as well as industry-based researchers in various fields.
In the context of DL frameworks, it is pertinent to mention that
TensorFlow is the most popular DL framework and has seen widespread
adoption. Today, more than 1,600 people have contributed to the TensorFlow
GitHub repository~\cite{tf-github} and several hundred research papers have
utilized TensorFlow for both research and commercial applications. However,
TensorFlow in its early days was criticized for its slower
performance~\cite{dlbench} as well as lack of support for efficient
distributed training and limited support for High Performance Computing
(HPC) systems. To this end, recent efforts by the TF developers as well as
the open source community are commendable and performance has significantly
improved for both single-GPU/single-node as well as multi-GPU/multi-node
(distributed) training. The gRPC~\cite{grpc} library, which is the official
distributed training infrastructure for TF, has been optimized for tensor
transfers (fewer memory operations), but still uses the relatively slow standard
Ethernet networks. However, gRPC can take advantage of InfiniBand (IB) using the IP
over IB (IPoIB) protocol which offers significantly better performance. At the
same time, the community has been actively exploring Message Passing Interface
(MPI) -- a de facto standard for the HPC community -- based designs to improve
distributed training on HPC clusters. 
However, the active interest and contributions from the community have led to
several disjoint efforts and fragmentation in the way users can take advantage
of the advanced distributed training designs in TF. The two broad challenges
that we investigate in this paper are: \textit{"1) What is the most efficient
tensor communication (for gradient aggregation) framework for TensorFlow and 2)
How can we improve the performance of this communication using CUDA-Aware MPI?}
Several detailed questions follow
this broad challenge. It is pertinent to note that little in existing
literature can offer insights to the following key questions, which are of
significant interest if large-scale distributed training is employed. 
\begin{itemize}
\item What are the available choices for distributed training using TensorFlow
for a given execution platform?
\item What are the key features and performance characteristics of the various
distributed training approaches for TensorFlow?
\item How can we optimize the performance of distributed training using
TensorFlow on modern HPC systems?
\end{itemize}
\MySubsection{Contributions}
To the best of our knowledge, this is the first paper that offers a
comprehensive landscape that highlights, evaluates, and optimizes a
diverse set of approaches that deal with distributed DNN training using
TensorFlow at scale. In this paper, we make the following key contributions:
\begin{itemize}
\item We provide an in-depth categorization, design analysis, and performance
characterization of \textit{distributed training approaches for TensorFlow}
using state-of-the-art DNNs like ResNet-50, MobileNet, and NASNet.
\item We propose a truly CUDA-Aware MPI\_Allreduce design that exploits 1) CUDA
kernels to perform large reductions on the GPU and 2) A pointer cache to avoid
overheads involved in queries to the CUDA driver. 
\item We illustrate benefits of the proposed MPI\_Allreduce
optimizations using micro-benchmarks as well as application workloads
(tf\_cnn\_benchmarks) using TensorFlow and Horovod.
\item We present a comprehensive and large-scale performance evaluation (up to
128 GPUs) for all the available distributed training approaches on three modern
HPC systems. 
\end{itemize}
\MySection{Overview of Communication Libraries for Distributed Training: Past, Present, and Future}
\label{sec:background} 
We first provide a brief historical perspective on various DL frameworks and how
they utilize communication libraries for distributed training. Next, we describe
communication libraries for distributed training using TensorFlow. We conclude
this section with a discussion on how TensorFlow is moving towards a different
system to handle community contributions and how it may affect the distributed
training approaches we discuss in Section~\ref{sec:analysis}. 
\MySubsection{DL Frameworks and Communication Libraries}
\label{sec:comm-libs}
Most ML/DL frameworks started with single-node/single-GPU designs. Caffe, for
example, had no support for multi-node distributed training and only external
efforts~\cite{scaffe,intel-caffe,firecaffe} provide such support. However, the
exponential growth in the size of DNN architectures and an ever-increasing need
for speed has forced the ML/DL framework designers to rethink their strategy
and to start utilizing existing communication schemes or design their own
libraries from scratch. Microsoft Cognitive Toolkit (CNTK)~\cite{cntk} is based
on an MPI design whereas Caffe2~\cite{caffe2} uses the Gloo~\cite{gloo}
collective communication library developed by Facebook, which is similar to
NVIDIA's NCCL~\cite{nccl} library. Gloo exploits the InfiniBand verbs interface
and Remote Direct Memory Access (RDMA) technology to offer various
reduction algorithms.
Apart from the communication libraries that come integrated with the frameworks,
there are external (commercial) efforts to enable efficient distributed
training. For example, IBM has developed the PowerAI Distributed Deep-learning
Library (DDL), which uses a multi-ring reduction algorithm. The library is built
on top of IBM's Spectrum MPI (a derivative of OpenMPI) and as such supports all the
network interfaces that are supported by the OpenMPI library. According to IBM, 
the library can be integrated with TensorFlow, Caffe and Torch. Intel has developed
the Intel Machine Learning Scaling Library (MLSL)~\cite{intel-sysml}. This
library is built on top of Intel MPI and therefore supports various
interconnects such as InfiniBand, Omni-Path, and Ethernet. The library offers a
set of communication primitives that neural network frameworks can take
advantage of when performing distributed communication. According to the
research paper it is integrated with Intel-Caffe, TensorFlow, and Intel's own
neural-net compiler called nGraph. To bring support to TensorFlow, the library
modifies the Horovod~\cite{horovod} library by inserting the MLSL 
communication primitives.
\MySubsection{Communication Libraries for TensorFlow}
\label{sec:tf-comm-libs}
Our focus in this paper is on distributed training using TensorFlow, which, by
default, can be realized using the official Google RPC (gRPC) library. gRPC is a
generic point-to-point communication library and has no collective communication
support. It works in a simple client/server style fashion and has been tightly
integrated with TF. However, TF has been extended by contributors to take
advantage of additional communication libraries like MPI, RDMA Verbs, and NCCL.
We discuss the libraries that TF uses in more detail in the following sections. 
\vspace{0.5ex}
\noindent \textbf{gRPC} 
is a point-to-point RPC library which exchanges data using
a pre-defined message format described in the protobuf
language~\cite{protobuf}. Each node launches a server process 
responsible for receiving messages from the other nodes. At the launch of TF,
the user has to specify IP and port number of the listening server for all the other nodes in the
training run. gRPC offers no CUDA-Aware operations, and as such, all data is first
staged on the host before being sent over the network. Recent optimizations have
however reduced the number of copy operations required to convert GPU buffers
into protobuf messages. All the communication operations are handled by a group
of threads which allow overlapping data transfers for optimal performance. gRPC
uses the HTTP/2 protocol on top of TCP/IP based networks. The generality of the
library causes it to also be widely used outside of the deep
learning application area.
\vspace{0.5ex}
\noindent \textbf{Message Passing Interface (MPI)} is a de facto standard for
expressing distributed-memory programs. Several implementations of the MPI
standard like MPICH~\cite{mpich}, MVAPICH~\cite{mvapich2-url},
OpenMPI~\cite{openmpi}, and CrayMPI~\cite{craympi} have been developed and
optimized over the period of several years for various processor architectures and high-performance interconnects like %
High-speed Ethernet (HSE) and InfiniBand. In recent years,
accelerators like NVIDIA GPUs have been adopted by most HPC
systems~\cite{top500}. As a result, MPI extensions have been proposed
to support efficient communication between GPUs. Initially, without
the capability of direct access of GPU memory, MPI applications required
explicit copying of GPU data to a staging buffer in host memory in order to push
the data to the network. Similarly, data movement from CPU to GPU was needed
after receiving the data on the host through an MPI\_Recv operation. This
significantly impacts performance as well as productivity. Thus, several MPI
libraries including OpenMPI and MVAPICH2-GDR~\cite{mvapich2-url}
provide \textit{CUDA-Aware} MPI primitives to transparently perform such copy
operations. These CUDA-Aware MPI libraries significantly improve performance and
productivity for MPI+CUDA applications.
\vspace{0.5ex}
\noindent \textbf{NVIDIA NCCL} offers multi-GPU communication primitives. NCCL
(pronounced Nickel) has been developed by NVIDIA to efficiently tackle
communication for DL workloads. NCCL's API closely resembles the MPI interface
and provides communication primitives for broadcast, all-gather, reduce,
reduce-scatter, and all-reduce. Precisely, NCCL's goal is to provide fast
communication of messages between GPUs in dense multi-GPU machines like the
DGX-1 and DGX-2~\cite{dgx2}. NCCL 1.x was, in fact, a single-node library.
However, since its introduction, NCCL has significantly evolved and now offers
multi-node collective operations using IB verbs. To launch NCCL2 applications,
users still need to rely on an out-of-band mechanism for connection management.
MPI launchers like ``mpirun'' are used to set up connections and assign ranks
while NCCL primitives are used for the actual communication.
\MySubsection{Future TensorFlow Developments}
\label{sec:bgnd-future-tf}
Google is currently redesigning how contributions are included in the TensorFlow
source tree. When TensorFlow 2.0 is released, a number of options that we have
analyzed in this paper might not be available anymore. However, Google
understands the importance of supporting multiple networking options and is
actively working on creating a special interest group (SIG) related to
networking. At the moment of writing, the APIs to add additional communication
methods to TF have not been formalized. However, there are plans
to re-enable the current contributions and introduce support for new
communication APIs such as UCX~\cite{ucx}.
\MySection{Analysis of Design and Performance: Several Distributed Training Approaches for TensorFlow} 
\label{sec:analysis}
We now dive into the various distributed training designs for TensorFlow and
group them into categories based on how the communication libraries are
exploited. Next, we provide the performance analysis for all of these
approaches. A high-level hierarchy of all these TF approaches is illustrated in
Figure~\ref{fig:overview}.
\MySubsection{gRPC-based Designs}
\label{sec:ps-approach}
The standard distributed training method for TensorFlow is the so-called
Parameter Server (PS) model~\cite{Li:2014:SDM:2685048.2685095}. 
In this model, there are worker processes that perform the heavy compute work and separate parameter
server processes responsible for combining the worker processes results. 
In other words, the PSs are responsible for storing and updating the
parameters (e.g., weights) of the neural network. TensorFlow programs can specify
the number of workers and number of PS tasks that should be used. The PS tasks
generally do not require much computing power and can run on nodes without
accelerators. Depending on the nodes used it is possible to run both a worker
process and a PS process on the same machine. TensorFlow comes with a set of
support classes that are built on top of the worker and PS instances to enable
monitoring and restarting the training process. This can be used for
checkpointing (saving) the training state or for fault tolerance in case a
worker node crashes due to hardware failure. TensorFlow's PS model utilizes the gRPC
library for communication between processes. It allows programs (optionally
written in different programming languages) to communicate via the TCP/IP
protocol by using RPCs and uniform data buffers. However, the
user is responsible for configuring the end-points for each of the launched
processes. This can be a labor-intensive task and requires knowledge about the
cluster and used nodes. TensorFlow uses the following steps to exchange tensor
buffers between the processes:
\noindent The {\tt producer} process:
\begin{enumerate}
  \item After a tensor, for which it has been determined that it has to be sent to another process,
  has been computed it is placed on a table.    
  \item If there is no outstanding request for the tensor then it stays in the table until
  a request is received. 
  \item If there is a request for this tensor then it is served immediately and removed from the waiting table.  
\end{enumerate}
\noindent The {\tt consumer} process:
\begin{enumerate}
\item Send a tensor request to the producer.
\item Wait until the producer returns the requested data.
\item Once the data has been received continue processing of the tensor graph
\end{enumerate}
 
\begin{figure}[htbp]
\centering
\includegraphics[width=0.8\columnwidth]{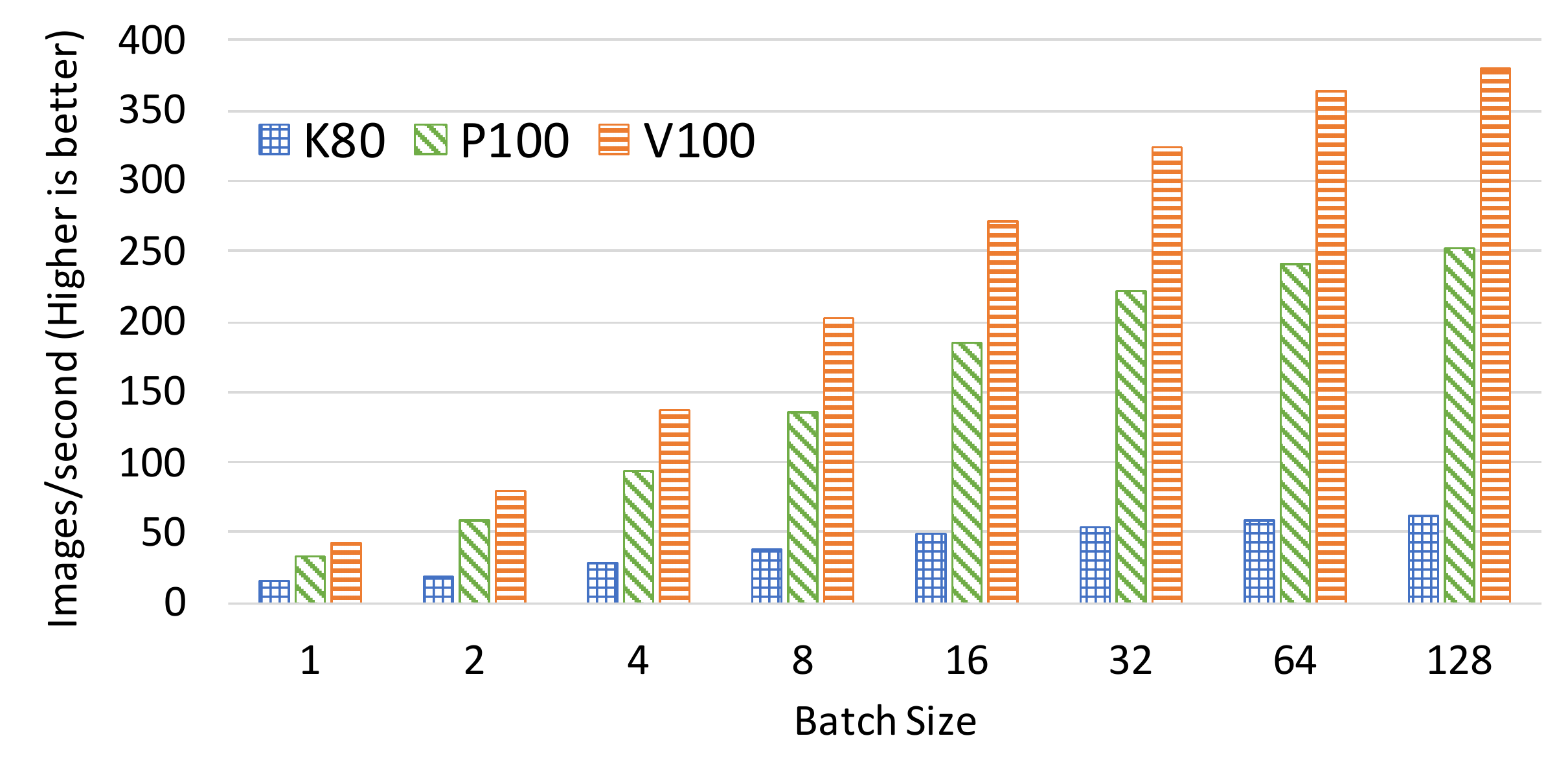}
    \vspace{-1\baselineskip}
\caption{Effect of Batch Size on Performance for Different GPU Architectures}
\label{fig:batchsize}
\end{figure}

\begin{figure*}[htbp]
    \centering
    \includegraphics[width=0.7\textwidth]{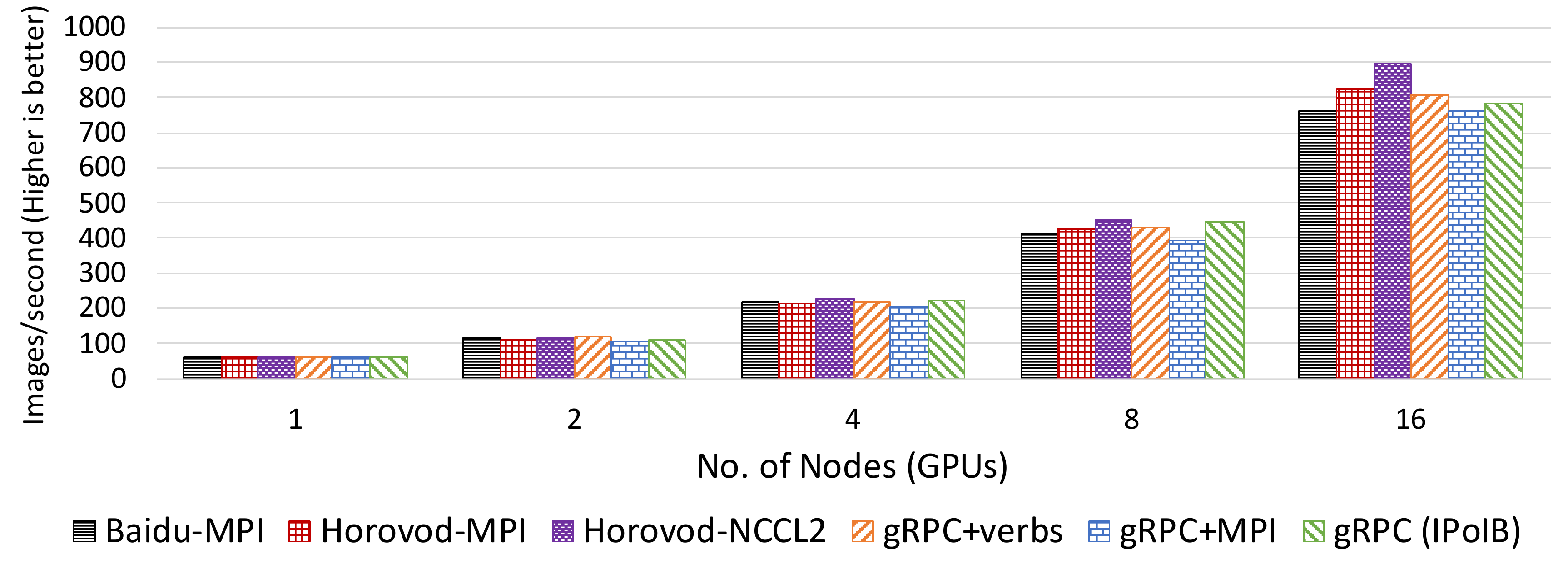}
    \vspace{-1\baselineskip}
    \caption{Performance Analysis of Six TensorFlow Approaches for Distributed
Training of ResNet-50 on the RI2 Cluster (see Section~\ref{sec:ri2}). 1) gRPC was executed
using IP addresses of the InfiniBand interface to make sure that tensor
communication happens using the efficient IPoIB channel. 2) MPI refers to the
MVAPICH2~\cite{mvapich2-url} MPI library for all cases. 3) The latest version of
NCCL2 (2.3.4) was used for all the experiments.} 
\label{fig:resnet50-ri2}
\end{figure*}

\begin{figure}[htbp]
    \centering
    \vspace{-1\baselineskip}
    \includegraphics[width=0.8\columnwidth]{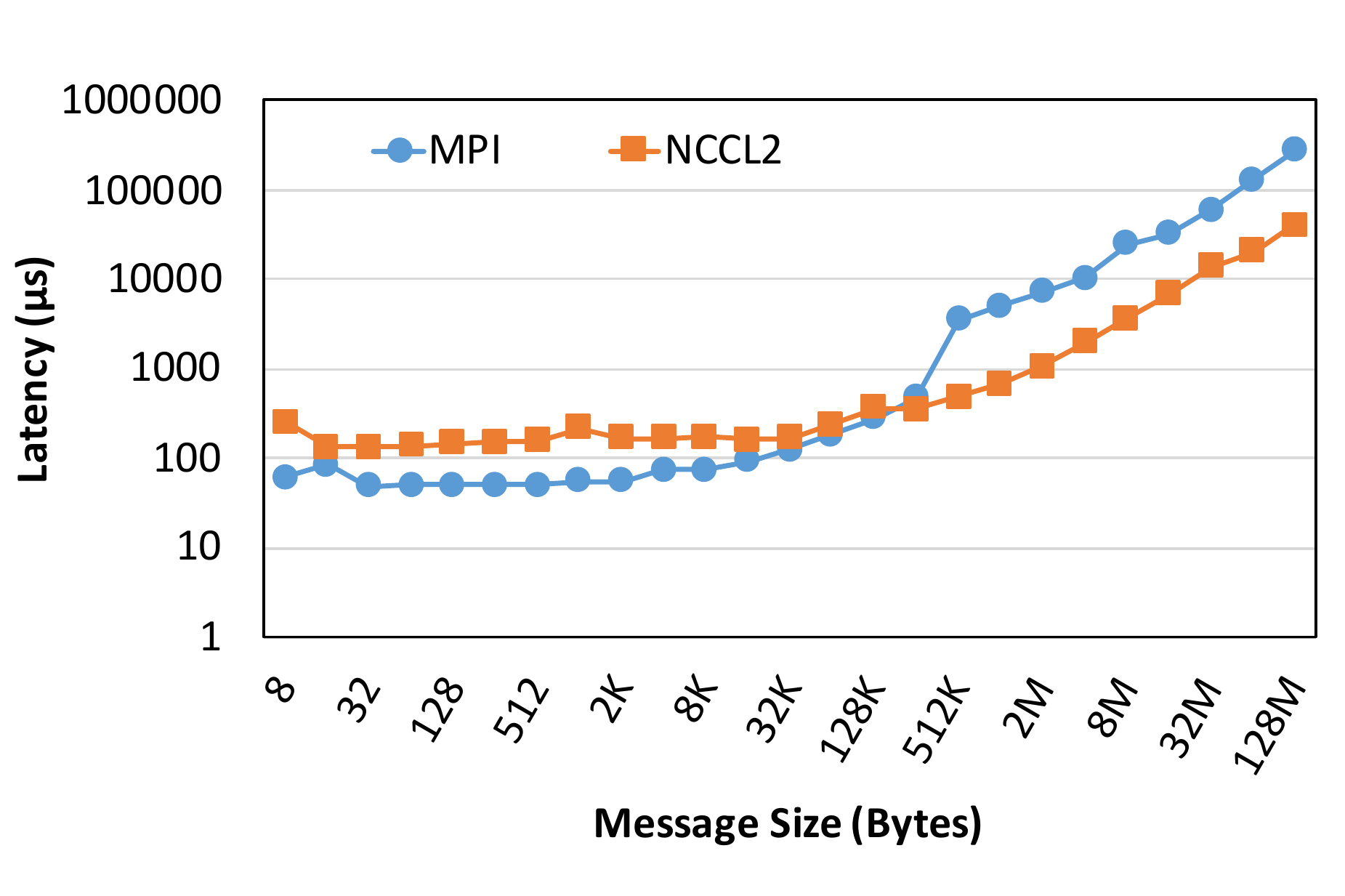}
    \vspace{-1\baselineskip}
    \caption{Micro-benchmark Performance for MPI\_Allreduce and NCCL Allreduce
        on the RI2 Cluster (see Section~\ref{sec:ri2}). 1) MPI refers to existing
MVAPICH2 MPI Library and 2) NCCL2 refers to NCCL 2.3.4}
\label{fig:allreduce-ri2}
\end{figure}

\noindent This is an example of a {\tt pull model} where the data is pulled from producer to
consumer. The gRPC library will send the tensor request message and the return
message that contains the tensor data. This data is encoded via the protobuffer 
library~\cite{protobuf}. If the tensor was computed on the GPU, then the tensor
data first has to be completely copied to the host memory before it can be
encoded in the protobuffer format and transferred. If the tensor's
destination is in GPU memory, then similar steps will have to be taken, where the
decoding takes place in the host memory before the decoded data is copied to the
GPU.
\vspace{0.5ex}
\noindent \textbf{Challenges in Extending gRPC-based Designs:} The tight
integration of TF with gRPC makes it non-trivial to add support for additional
network protocols. However, TensorFlow's design allows for offloading just the
tensor transfers, which are the most data-intensive communication operations
during the processing of neural networks. With the high-performance GPUs (and
other accelerators) currently available it is critical that the data transfers
happen as fast as possible, otherwise, the GPUs would sit idle waiting for data.
The more administrative communication operations such as setting up the network
and controlling the execution are less time critical and will always be
performed via the gRPC stack. Hence, no matter if the tensor communication is
offloaded to another method, the requirement for setting up communication
endpoints stays in place. This offloading functionality is what enables the
RDMA over Converged Ethernet (RoCE) support that Google's internal TensorFlow
version supports. It also enables (external) contributors to add support for
other network stacks, see the next section.
Because the changes happen on the
deeper levels of TF, the user does not have to make any changes to his designed
neural net. The only change required is the specification of the communication
protocol in the execution script. However, to take advantage of these
contributed packages the user will have to build TF from source and point the
build process to the correct libraries. gRPC-based communication can also
transparently take advantage of the IPoIB~\cite{ipoib} interface if 
IB is the underlying interconnect.
\MySubsection{gRPC+`X' Designs}
\label{sec:grpcx-design}
We broadly discuss three gRPC extensions: 1) gRPC+MPI, 2) gRPC+Verbs, and 3) gRPC+GDR.
\vspace{0.5ex}
\MySubsubsection{MPI}
\noindent The gRPC+MPI method adds support for sending tensors using MPI primitives
(see Section~\ref{sec:tf-comm-libs}). Depending on the support that the MPI library offers, this
allows TF to make use of advanced network interconnects such as IB and
Omni-Path. The current implementation of the gRPC+MPI method uses a single thread
for all the MPI related operations. On one side, this increases compatibility
with MPI frameworks that are not thread-safe, but on the other side, it can
hamper performance. Especially when many small data tensors are exchanged there
would be enough bandwidth available for parallel transfers which could increase
overall performance. MPI processes are uniquely identified using
their rank, which is incompatible with the TCP/IP connection strings that gRPC
uses to identify processes. Therefore a mapping is created from the gRPC
identifiers to the MPI process ranks during the initialization of TF.
\MySubsubsection{Verbs}
The gRPC+Verbs protocol uses the RDMA-verbs API. By directly using the verbs
APIs, there is no need for setting up a global execution environment as
required by MPI. However, many features that are available in MPI have to be
manually implemented. Examples include features like pinned memory buffers to
improve the performance of data-transfers, and the use of GPUDirect RDMA. The
pinned memory functionality is implemented in the
gRPC+Verbs~\cite{yahoo-tf-verbs} protocol and can be utilized via the tf.contrib
package. Another approach to exploit IB Verbs for gRPC in a more transparent
fashion has been presented in~\cite{Biswas:HiPC18}. We refer to this as
Accelerated gRPC in Figure~\ref{fig:overview}. 
\MySubsubsection{GDR}
The last contributed protocol is the gRPC+GDR protocol. This protocol adds
GPUDirect RDMA (GDR) functionality for NVIDIA GPUs. This allows tensors to be directly
read/written from/to the GPU memory to/from the network adapter without having to first transfer data
to the host memory. This saves a full memory copy and reduces latency.  In order
to take advantage of this protocol, it is required that the system used has the
correct drivers and hardware installed to support this operation. Currently, it
is only supported for Intel and AMD CPUs combined with an NVIDIA GPU and
Mellanox IB Host Controller Adapter (HCA). The details of this extension are
available in~\cite{grpc-gdr-paper}. It is pertinent to mention that GDR support
has been exploited by MPI runtimes like MVAPICH2 and OpenMPI %
and can be utilized transparently on systems that support this feature and have
the required GPU/HCA hardware available. Unfortunately, gRPC+GDR designs
did not run properly on any of our clusters, so no results or discussion is
provided for this design. 
\MySubsection{No-gRPC Designs} 
\label{sec:reduce-approach}
Besides the parameter server method to combine (reduce) the results of worker
processes, there is the option to use collective operators to perform the
reduction. When using these collective operators, there are no separate
processes responsible for gathering and updating the network parameters. 
Instead, the task is distributed over the worker processes. The collective operator
required for these operations is called Allreduce; a communication primitive
widely used in the HPC/MPI community. 
There are two `No-gRPC' designs available as contributed packages to TF: 1)
Baidu and 2) Horovod. Both introduce the so-called reduction operators for
TF~\cite{tf-ops} and exploit the Allreduce primitive. Because the reduction
based approach is not available in the standard TF version, it requires a bit
more work to make use of it than the protocol changes described above. The way
these methods are integrated is by overloading a subset of the default TF
operators. During execution, these packages get invoked as part of the
network (model) configuration phase and detect which of the parameters used in
the execution graph need to be combined. Next, additional reduction operators
are inserted into the execution graph. During the execution of the graph, these
operators are executed, which invokes the corresponding implementation of the
Allreduce operator.
A major advantage for the users of these reduction based methods is that the
steps required to setup the execution of the neural network are much simpler. It
is no longer required to launch and configure the additional parameter servers.
Furthermore, because both these methods are based on the MPI execution model,
the user does not need to configure the endpoints explicitly. Instead, the MPI
library/launcher can handle the discovery and connection management. 
\MySubsubsection{Baidu}
Baidu's design is part of the tf.contrib package in the TF codebase. It is
called tf.contrib.mpi\_collectives and can be enabled during the compilation and
configuration steps when building TF from source. This extension is not
currently available in the TensorFlow wheels (binary packages) available from
Google's website. Baidu's design contains a custom ring-reduce implementation of
Allreduce built on top of MPI\_Send and MPI\_Irecv primitives. This gives
access to the same MPI advantages as mentioned above, namely extensive network
support and support for CUDA-Aware MPI operators which reduces the number of
required memory operations.
\MySubsubsection{Horovod}
The second method, introduced by Uber, is available as an external Python
package called Horovod and can be installed using the {\tt pip} package
manager. The design of Horovod is based on the mpi\_collectives implementation
and as such the user side modifications are similar for these packages. However,
Horovod design offers some advantages of Baidu's design. 
Horovod can take advantage of two different reduction implementations. The first
is based on the standard MPI\_Allreduce operator. The reduction method
that this operator uses depends on the underlying MPI implementation. For
example, MVAPICH2-GDR now supports efficient CUDA-Aware MPI\_Allreduce designs as
discussed in Section~\ref{sec:opt-mpi-gdr} while default MPICH~\cite{mpich}
and OpenMPI only provide naive implementations of MPI\_Allreduce for GPU
buffers. More specifically, the actual reduction implementation in an MPI
runtime can differ based on the supercomputer or cluster that is used. Many
system vendors add custom communication methods that are designed and tuned
specifically for the specific hardware and network topology. The second Horovod
implementation exploits NVIDIA NCCL library, which is based on InfiniBand verbs
and CUDA kernels to handle communication across and within nodes, respectively.
NCCL2 currently cannot be used on the Piz Daint system because it has the
proprietary Aries interconnect, which does not support IB verbs. The MPI variant of
Horovod, on the other hand, is portable and can be used on the Piz Daint system
via Cray's CUDA-Aware MPI implementation.  
Horovod also provides a unique feature called ``Tensor Fusion''. When using this
method, several small tensors are combined in a single reduction operation. The
idea is that by performing several large reductions instead of many small ones
the latency is reduced and bandwidth is used more efficiently. The tensor fusion
feature is controlled via a runtime threshold parameter, and we experimentally
determine the best threshold for a given
platform. 
\MySubsection{Performance Analysis}
We now present a comprehensive performance comparison of all the aforementioned
approaches to perform distributed training using TensorFlow. In order to
understand the trends for distributed training, the most fundamental question
that often gets missed is: \textit{What is the best batch size one should use
for distributed training?}. In this context, the general intuition is that a
bigger batch size will lead to better performance. However, we perform
single-GPU experiments to understand this behavior in a more pragmatic fashion. 
Figure~\ref{fig:batchsize} shows the impact of the batch size for three GPU
generations: 1) Kepler (K80), 2) Pascal (P100), and 3) Volta (V100). The key
insight here is: ~\textit{faster GPUs offer better performance for larger batch
sizes up to the point of diminishing returns}. For ResNet-50, the sweet spot
seems to be 64 for all three GPUs. Thus, we choose 64 as the single-GPU
batch-size baseline and utilize this for all distributed training experiments
performed for ResNet-50.
The performance numbers for all six approaches except
grpc+gdr are presented in Figure~\ref{fig:resnet50-ri2}. The main insight we
gain from these experiments is: \textit{gRPC and grpc+`X' designs are, in
general, slower than Horovod designs}. Secondary observations include: 1) Baidu's
design despite using the ring-allreduce lags behind Horovod-NCCL and grpc+`X'
designs and 2) Horovod-MV2 is always slower than Horovod-NCCL2. We have used
TensorFlow version 1.10.0 for all our experiments. 
\vspace{0.5ex}
\noindent \textbf{Performance of Allreduce:} The observation about the different
performance we get for NCCL2 compared to MVAPICH2 for Horovod designs merits
further investigation. This is because the only difference in Horovod-NCCL and
Horovod-MPI design is the utilization of the corresponding Allreduce primitive.
Thus, we perform additional experiments using MPI/NCCL benchmarks to better
understand this behavior. Indeed, the performance of NCCL Allreduce is better
than MPI\_Allreduce in MVAPICH2, as shown in Figure~\ref{fig:allreduce-ri2}.
We performed a thorough analysis of Allreduce designs in the MVAPICH2 library
and found a clear opportunity for performance optimizations to deal with DL
workloads (large message sizes). Based on this, we propose optimizations for
Allreduce and highlight the performance benefits we observed for these
enhancements in Section~\ref{sec:opt-mpi-gdr}.
\MySection{Optimizing \textit{tf\_cnn\_benchmarks} for HPC Systems}
\label{sec:benchmark-script}
The performance analysis presented in Section~\ref{sec:analysis} was made
possible via the official TensorFlow benchmarking scripts called 
tf\_cnn\_benchmarks~\cite{tf-cnn-bench}. To evaluate various platforms and
distributed TF implementations, we modified these application level benchmarks
and added support for running Baidu's design. We also modified the scripts to
enable gRPC and gRPC+X runs using SLURM~\cite{slurm}. During the start-up phase,
we pull in the SLURM environment variables in order to determine the total
number of launched benchmark instances and their unique IDs (rank). This is
required to setup the configurations that do not use MPI, but also require that
each process is uniquely identified. The unique ID is consequently used to
determine the type of process (worker, or parameter server) and to setup the
underlying network connections. Because this is based on the SLURM environment
variables it is trivial to adapt this to other workload managers, by using
the variables that are specific to that workload manager or particular {\tt
mpirun} version. Once this setup is complete, and the various processes are
connected to each other, the neural network will be initialized. 
This modification to tf\_cnn\_benchmarks allowed us to test all of the several
distributed training configurations with exactly the same neural network design
and data processing routines. This ensures that any performance differences
between the various networking options can be directly attributed to the
reduction algorithm and the networking library. The benchmark suite allows
users to test the performance of various convolutional neural networks using
TensorFlow. For our tests, we use three different image classification networks,
MobileNet~\cite{DBLP:journals/corr/HowardZCKWWAA17},
ResNet-50~\cite{2015arXiv151203385H} and
NASNet-large~\cite{DBLP:journals/corr/ZophVSL17}. To prevent that our results
are influenced by file I/O (disk) performance, we only use synthetic input data.
The training phase for all the various approaches will remain common
and contains exactly the same operations as the real neural network training with
the difference being the usage of synthetic data instead of real images that
need to be read from a storage subsystem. After a number of warm-up iterations, a
set of ten iterations will be performed to determine the image throughput rate.
Note that because synthetic data is used we purely measure the GPU and network
performance for the multi-node cases but only the GPU performance for a single process case. 
The CPU (used for decoding image data) and storage media (used for loading data)
are not part of our tests since we focus only on the scaling characteristics of
distributed training. However, when doing real training runs these components
will influence the final performance, especially the storage media will be
important as a large number of GPUs will have to be fed with fresh training data
during each iteration.
\section{Optimizing CUDA-Aware MPI Primitives}
\label{sec:opt-mpi-gdr}
\begin{figure}[htbp]
    \centering
    \includegraphics[width=0.9\columnwidth]{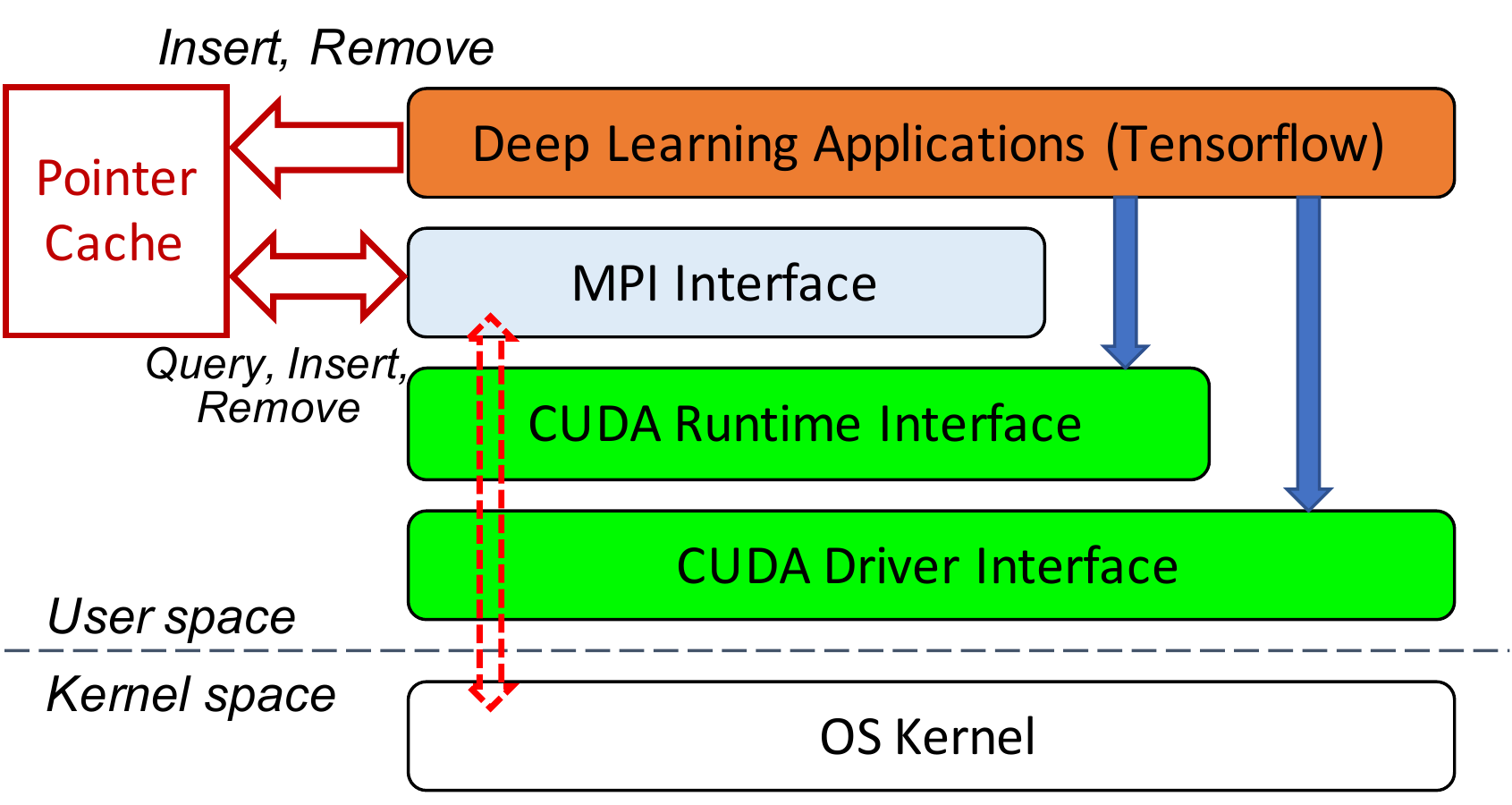}
    \caption{Pointer Cache Design (dark-red box and arrows) to avoid expensive queries to the CUDA driver (red dashed arrow).}
    \label{fig:pointer-cache}
\end{figure}

\begin{figure}[htbp]
    \centering
    \vspace{-1\baselineskip}
    \includegraphics[width=0.8\columnwidth]{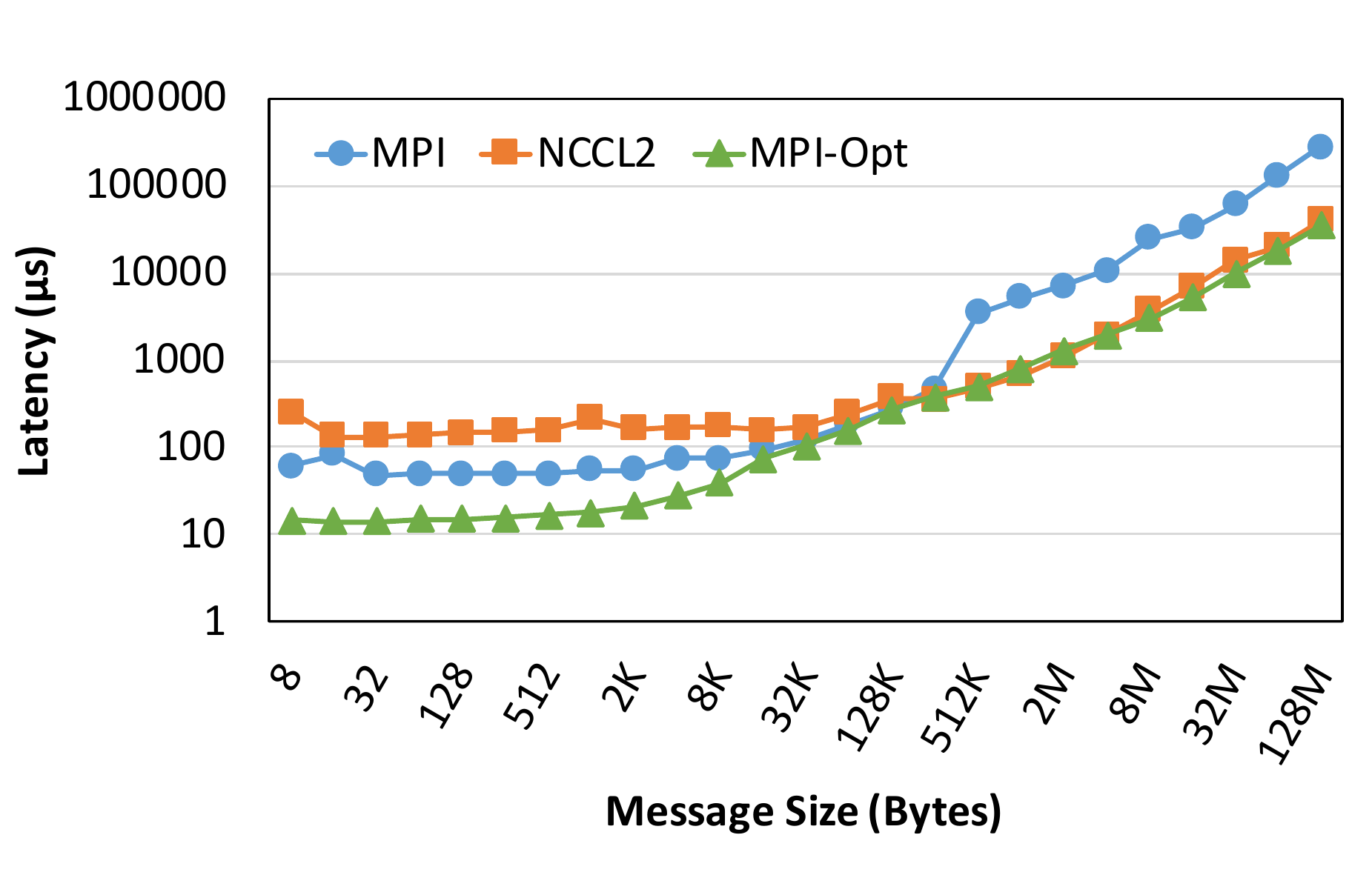}
    \vspace{-1\baselineskip}
    \caption{Benefits of the Allreduce optimizations. 1) MPI refers to
existing MVAPICH2 MPI Library. 2) MPI-Opt refers to the optimizations
made available in MVAPICH2-GDR 2.3rc1}
    \label{fig:impact-gdr-allred-opt}
\end{figure}

Based on the analysis in Section~\ref{sec:analysis}, we observed that
the MPI\_Allreduce primitive is a significant performance bottleneck for the
Horovod-MPI approach. Thus, the primary challenge we have to deal with is a better
and more efficient MPI\_Allreduce design to tackle DL workloads at scale. To
address this, we propose two major design optimizations: A) Truly CUDA-Aware
reduction method that exploits CUDA based computations in
addition to GPUDirect-based communication, and B) An advanced caching scheme
for GPU-based (device) pointers to avoid query overhead in the critical path. 
\MySubsection{CUDA Kernel-enabled Allreduce for large messages}
\label{sec:kernel-allreduce}
Although offloading reduction operations to GPU kernels has been discussed
before~\cite{Oden:CCGRID14, Chu:CCGRID16, Luo:HPDC18}, for large
messages it has not fully leveraged the GPU computing power for Allreduce
algorithms.  To perform large-message Allreduce operations efficiently, a
variety of ``reduce-scatter followed by allgather (RSA)'' algorithms are widely
used. Two popular algorithms are: 1) Ring-based
RSA~\cite{Patarasuk:2009:JPDC}: first a virtual ring is constructed between
processes. Next, each process sends a chunk of data to its left neighbor and
performs reduction upon the chunk received from its right neighbor for $(p-1)$
steps, where $p$ denotes the number of processes. At the end of this
reduce-scatter phase, each process owns a partial final result vector. Then
a similar ring-based allgather is performed to gather the complete final
result (notable implementations: NCCL and Baidu Allreduce). 2) Recursive
vector halving and doubling RSA~\cite{Thakur:2005:mpich-coll}: each process
pair exchanges half of the vector/message and reduces it. Next, the message
size is halved, and a different pair of processes is selected for $\log p$
steps. Allgather is performed in reverse and doubles the  message size at each
of the $\log p$ steps (notable implementations MPICH and MVAPICH2).  The first method
is popular because of the bandwidth-optimized feature, while the second method
is expected to be more efficient at scale. However, the existing MVAPICH2 implementation
of recursive vector halving and doubling RSA relies on the CPU to perform
reduction operations, which is a waste of GPU compute power.
In this paper, we propose and implement a CUDA-Aware MPI\_Allreduce primitive
using recursive vector halving and doubling RSA, with GPU kernel-enabled
reduction operations.  There are three major benefits: 1) significantly
reduces the reduction latency for large messages by fully utilizing
the massive parallelism and high-bandwidth memory of GPUs.  2) Avoids expensive
data movement from GPU to host memory. 3) Leverages advanced hardware and
software features such as GPUDirect RDMA to improve intra-node and inter-node
communication.
\MySubsection{Optimized Pointer Cache for Device Buffers}
\label{sec:pointer-cache}
The CUDA \textit{unified addressing} feature allows NVIDIA GPU devices to share
a unified address space with the host. This means that there is no distinction
between a device pointer and a host pointer, i.e., the same pointer value could
represent host memory or device memory at different times.  MPI libraries like
MVAPICH2 have implemented various algorithms to optimize MPI primitives for
host and device buffers. Therefore, the MPI runtime needs to identify the
buffer type before selecting the most performant algorithm.  The CUDA driver
provides a low-level API to perform this identification.  However, each MPI
call may need to query the buffer type multiple time even when the pointer
value remains unchanged. This incurs a significant delay in the MPI function as
the call accesses multiple driver modules as illustrated in
Figure~\ref{fig:pointer-cache} (red dashed arrow).
To mitigate this overhead, we designed a \textit{Pointer Cache}, which stores the pointer
information, as depicted in Figure~\ref{fig:pointer-cache} (dark-red solid box and arrows).
There are two ways to maintain the cache: 1) one-time driver lookup at MPI-level, 2) interception of 
the device allocation APIs at application-level. In the first approach, the MPI runtime caches 
(i.e., \textit{inserts}) the buffer type of the given pointer when it is seen for the first time. 
However, the runtime is not able to invalidate a cache entry when the buffer gets de-allocated 
by the application without notifying the MPI runtime. This leads to the second approach. 
We let the MPI runtime intercept CUDA memory management APIs, e.g., \textit{cuMalloc}
and \textit{cuFree}, and update the pointer cache accordingly. 
This way, the MPI runtime only has to query the cache but not maintain it.
This optimization improves the performance of all CUDA-aware MPI primitives, including the ones
primarily used by the deep learning frameworks.
\MySubsection{Performance Benefits}
To evaluate the proposed optimizations, we conducted experiments across 16
NVIDIA Tesla K80 GPU nodes at a local cluster (RI2).  We used micro-benchmarks
to compare the Allreduce performance between default MVAPICH2 ({\it MPI}),
NVIDIA NCCL ({\it NCCL2}), and the new MVAPICH2 implementation ({\it MPI-Opt}).
Figure~\ref{fig:impact-gdr-allred-opt} shows that the new method yields
4.1$\times$ faster Allreduce compared to the default MVAPICH2 solution for
small and medium messages (i.e., smaller than 128KB) due to the pointer cache.
Yet, {\it MPI-Opt} is 17$\times$ faster than {\it NCCL2} for Allreduce operation of
8-byte data.  For larger message sizes, the optimized GPU kernel-enabled
reduction scheme provides up to 8$\times$ (1.4$\times$) performance improvement
compared to the default MVAPICH2 ({\it NCCL2}) library. With the significant
benchmark-level improvements, we next present the application level performance
of the enhanced MPI primitives.
\MySection{Comprehensive Performance Comparison of Existing and Proposed Designs}
\label{sec:results}
We now present an in-depth performance comparison of all the distributed
training approaches with existing communication libraries as well as with our
proposed Allreduce design schemes. We perform various experiments on three
different clusters: 1) The RI2 cluster at The Ohio State University, 2) The
Owens cluster at Ohio Supercomputing Center (OSC), and 3) The Piz Daint cluster
at Swiss National Supercomputing Centre. 
\MySubsection{Performance of CPU vs. GPU based Training} 
 
\begin{figure}[htbp]
    \centering
    \includegraphics[width=0.8\columnwidth]{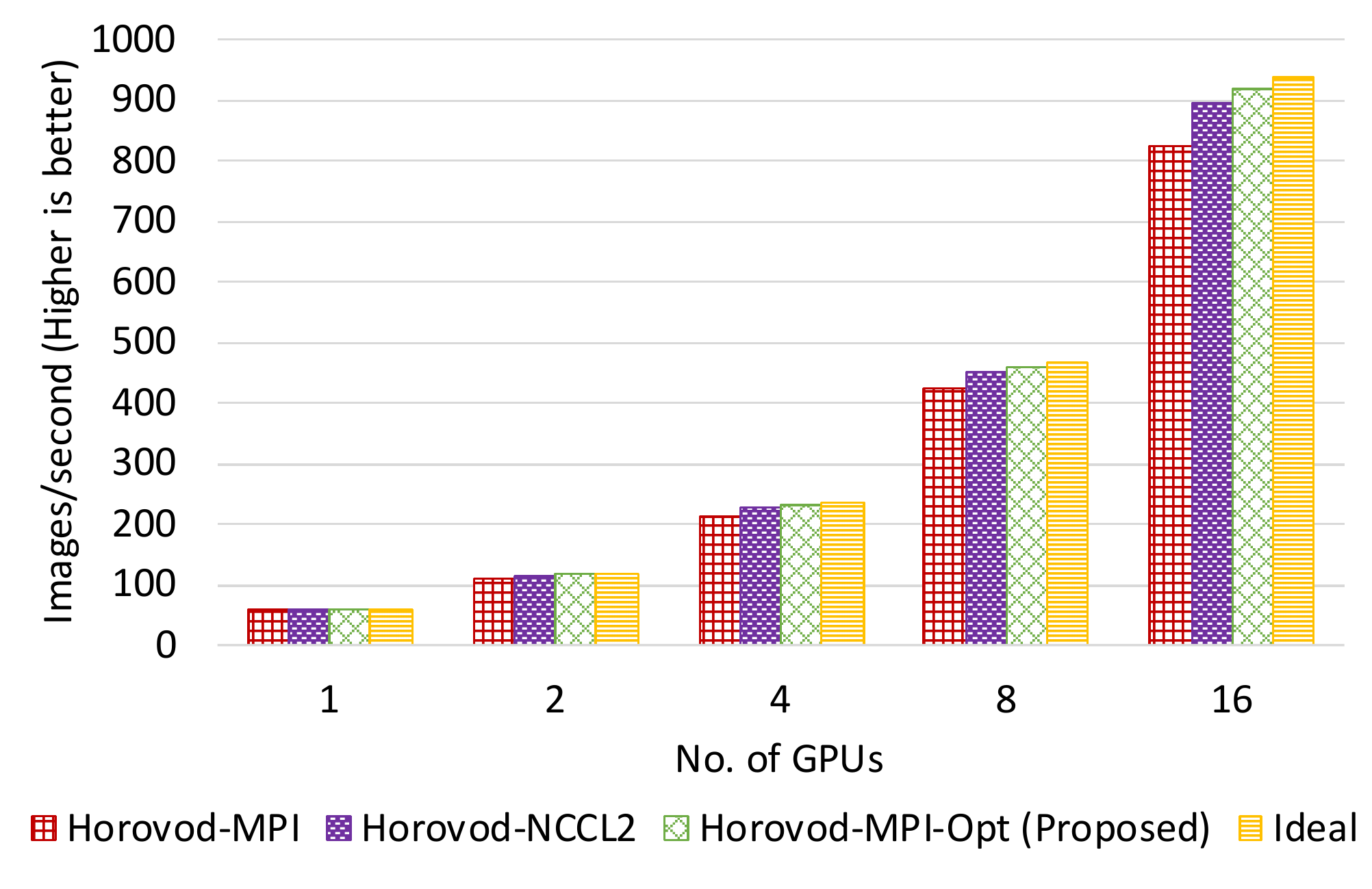}
    \vspace{-1\baselineskip}
    \caption{Performance comparison for ResNet-50 training using three Horovod
versions on the RI2 Cluster (up to 16 GPUs). 1) NCCL 2.3.4 was used for NCCL
experiments. 2) MPI refers to the MVAPICH2~\cite{mvapich2-url} library. 3) MPI-Opt
refers to the new Allreduce designs made available as part of the
MVAPICH2-GDR 2.3rc1 library.} 
\label{fig:resnet50-ri2-updated}
\end{figure}

\begin{figure}[htbp]
\centering
\includegraphics[width=0.8\columnwidth]{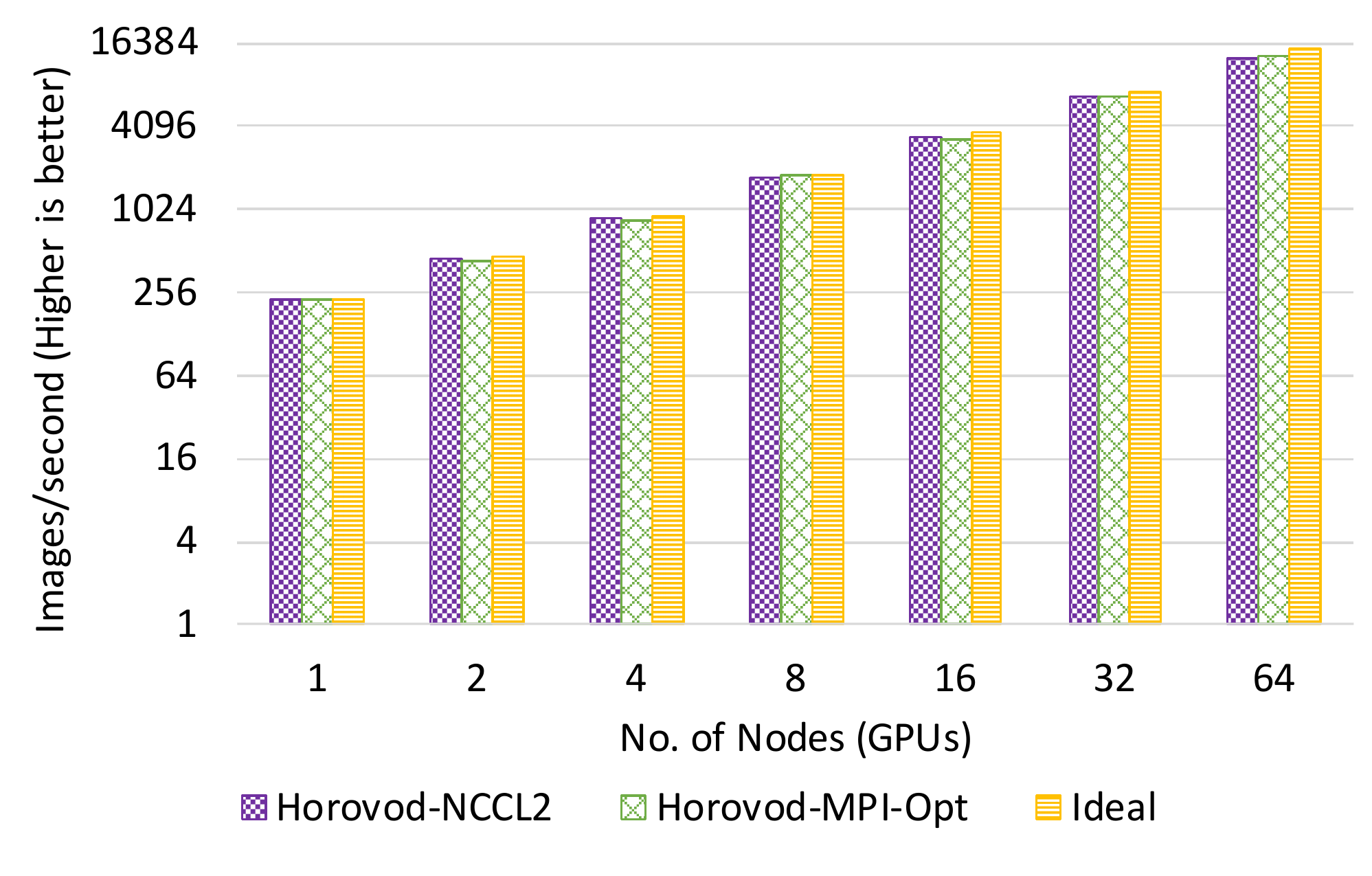}
    \vspace{-1\baselineskip}
\caption{Performance comparison for ResNet-50: Training performed using two Horovod
designs on the Owens Cluster (up to 64 GPUs). 1) NCCL 2.3.4 was used for NCCL
experiments. 2) Horovod-MPI-Opt refers to the design that takes advantage of the
new Allreduce implementation made available in the MVAPICH2-GDR 2.3rc1 library.}
\label{fig:owens-resnet}
\end{figure}

\begin{figure*}[htbp]
\centering
\subfloat[NASNet-large]{\includegraphics[width=0.35\linewidth]{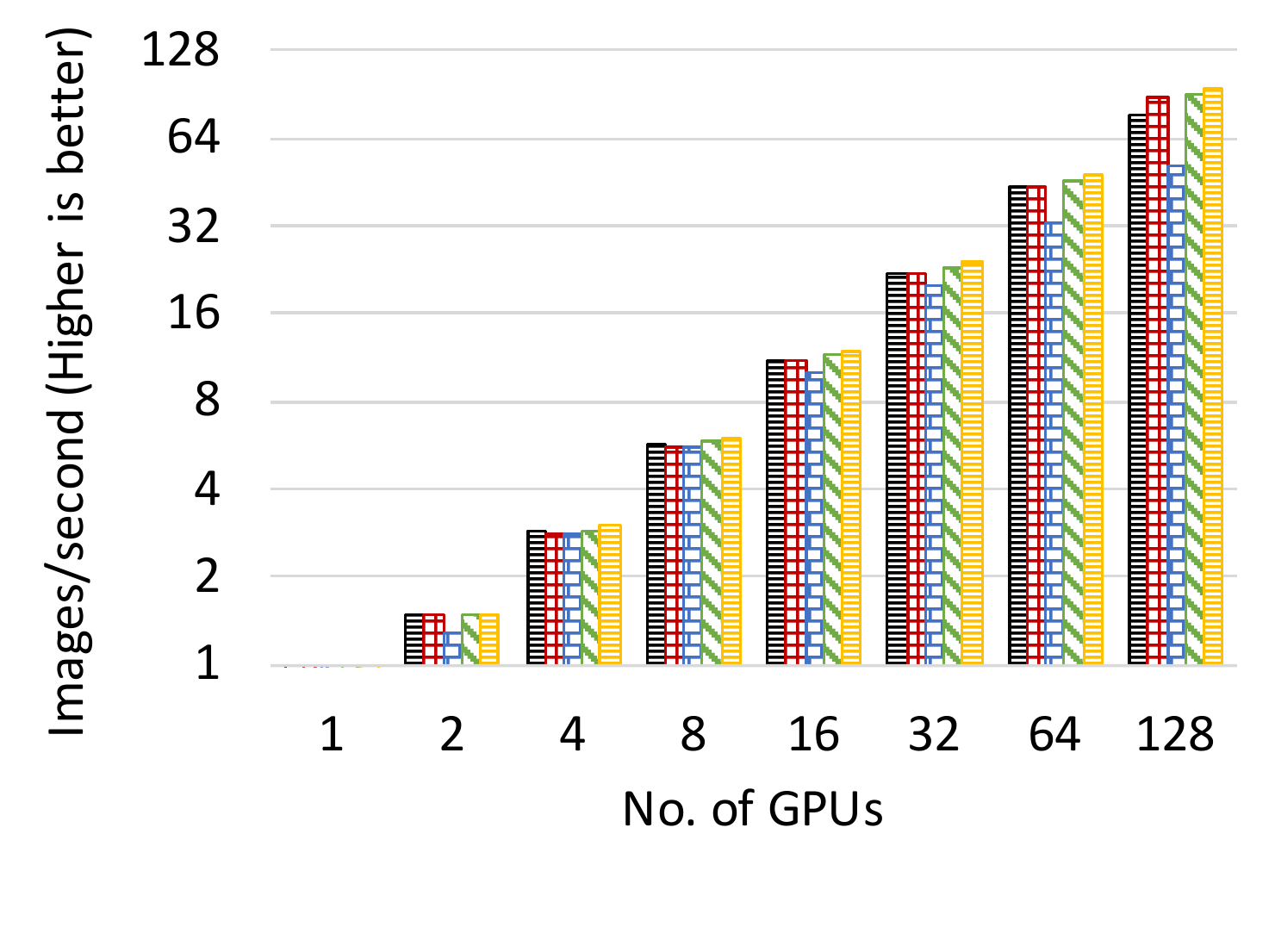}%
\label{fig:nasnet-pd}}
\hfil
\subfloat[ResNet-50]{\includegraphics[width=0.30\linewidth]{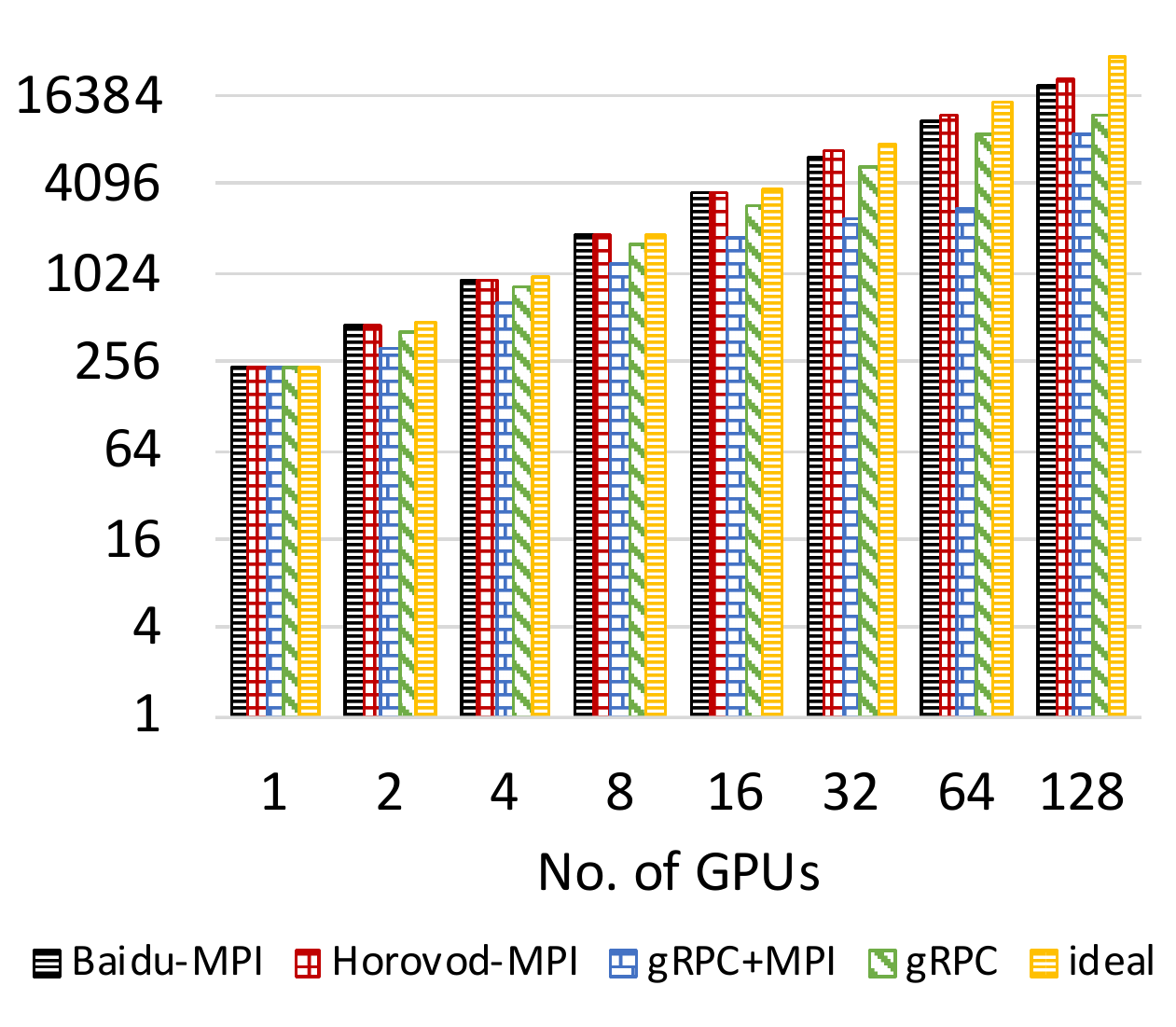}%
\label{fig:resnet-pd}}
\hfil
\subfloat[Mobilenet]{\includegraphics[width=0.30\linewidth]{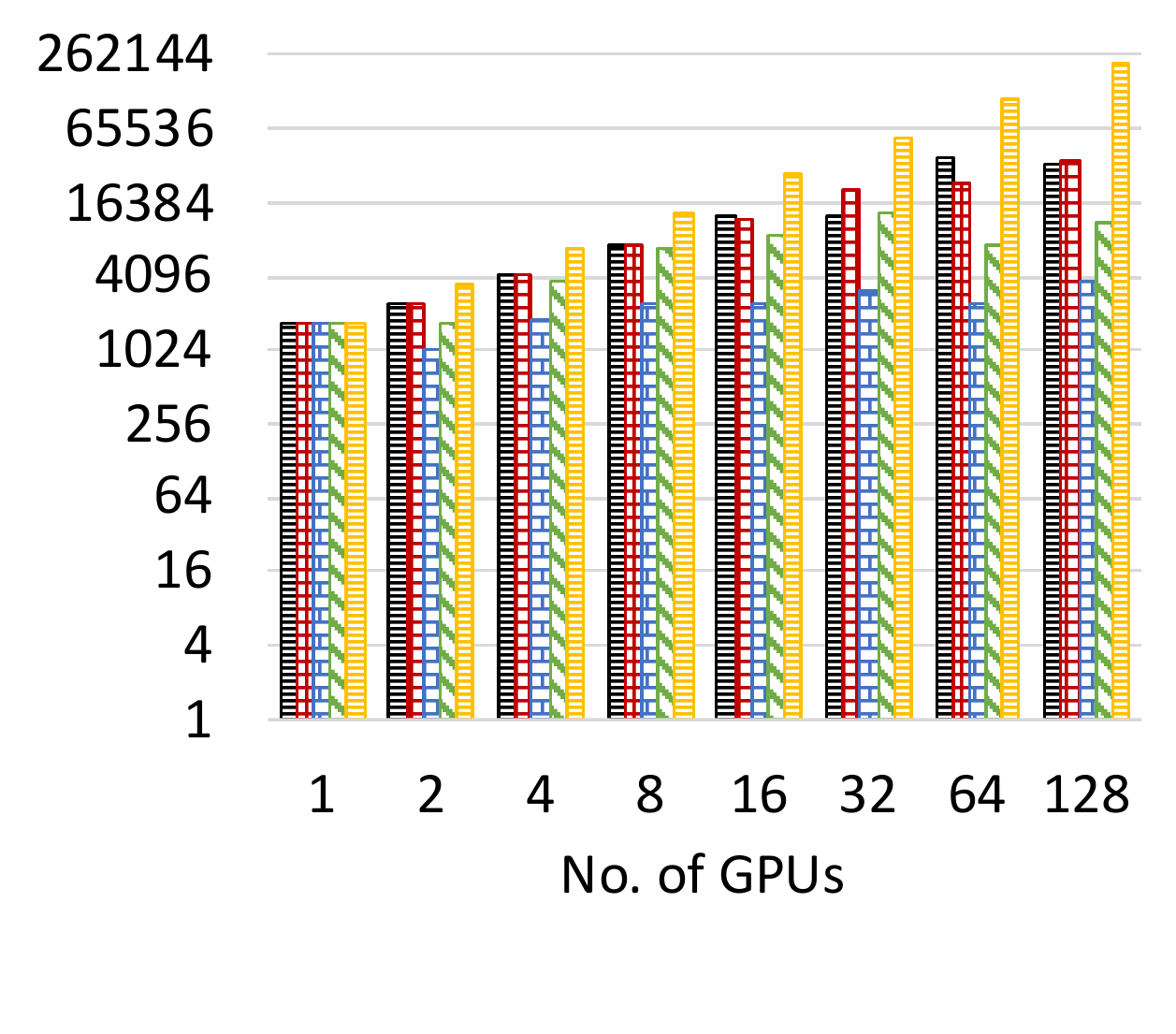}%
\label{fig:mobnet-pd}}
\caption{Performance Comparison for Various Distributed Training Approaches
using 1) NASNet-large, 2) ResNet-50, and 3) Mobilenet on the Piz Daint Cluster
(up to 128 GPUs). MPI refers to Cray-MPICH (v7.6) MPI Library.}
\label{fig:piz-daint-scale}
\end{figure*}

We performed several GPU and CPU based training experiments. However, there are
many additional vectors that need to be considered to provide a sound
performance evaluation. The default CPU installation of TensorFlow offers much
slower performance compared to GPU-based training. We also explored the
Intel-optimized TensorFlow version. However, we chose to only focus on GPU
results in this paper as the discussion and comparison of default CPU, optimized
CPU, and GPU based training is beyond the scope of this paper.
\MySubsection{RI2}
\label{sec:ri2}
\noindent The RI2 cluster at The Ohio State University consists of 20 nodes
connected via Mellanox SB7790 and SB7800 InfiniBand switches. Each node is
equipped with two 14-core Intel (Broadwell) Xeon E5-2680 v4 2.4 GHz processors,
128 GB DDR3 Memory, one NVIDIA Tesla K80 GPU, and one single port InfiniBand EDR
HCA. Based on the proposed optimizations for Allreduce discussed in
Section~\ref{sec:opt-mpi-gdr}, we now provide a comparison of the best performing
Horovod-NCCL approach (shown earlier in Figure~\ref{fig:resnet50-ri2}),
Horovod-MPI, and the proposed Horovod-MPI-Opt approach that takes advantage of
the optimized MPI\_Allreduce design. As we can clearly see in
Figure~\ref{fig:resnet50-ri2-updated}, Horovod-MPI-Opt performs much
better than Horovod-MPI as well as offering better/comparable performance as
Horovod-NCCL2 for most cases. 
The ``ideal'' bar for all the graphs in this section has been calculated using a
linear speedup formula: 
\noindent {\tt Ideal = Images/second (for 1 GPU) $\times$ \#GPUs}
\MySubsection{Owens}
\noindent Owens is a 23,392-core Dell Intel Xeon E5-2680 v4 machine with 160 GPU nodes
equipped with NVIDIA Pascal P100 GPUs. Each node is equipped with a dual-socket
Intel Xeon processor (28 cores) and an IB EDR HCA.
We have used the Owens cluster to extend our finding from RI2 to a larger scale
as well as a newer GPU generation. The cluster is heavily used by HPC
researchers across the US, so we limit our performance evaluation to only the
best performing TF variants from RI2. Figure~\ref{fig:owens-resnet} provides the
performance comparison of Horovod-NCCL2 and Horovod-MPI-Opt for the distributed
training of ResNet-50 on up to 64 Pascal GPUs. Clearly, we can observe that our
proposed Allreduce optimizations have enabled Horovod-MPI-Opt to achieve
near-ideal scaling that is better/comparable to Horovod-NCCL designs. It is
pertinent to mention that the Horovod designs have been the best for all
configurations on RI2 as well as the Owens cluster. 
\MySubsection{Piz Daint}
\noindent We now provide the scaling results up to an even larger scale (up to 128
Pascal GPUs) on the Piz Daint cluster. We also extend our evaluation for Piz Daint
with two more DNN architectures: 1)
MobileNet~\cite{DBLP:journals/corr/HowardZCKWWAA17} and 2)
NASNet~\cite{DBLP:journals/corr/ZophVSL17}. Each node of the Piz Daint cluster
is equipped with a single P100 GPU and a 12-core Intel Xeon E5-2690 v3 (Haswell)
CPU. The nodes are connected using the Cray Aries communication technology based
on the Dragonfly topology. The machine has thousands of compute nodes so the
actual placement (distance) between the processes is random and can influence
the actual execution time. The software stack is designed by Cray and optimized
for the machine. This means that we have limited control over the used (MPI)
libraries and just use what the system has available. Because of this software
limitation, we were not able to test out the Horovod-NCCL implementation on this
system as there is no support for IB verbs, which NCCL uses for inter-node
communication. For Horovod-MPI, we have used the default Cray-MPICH (v7.6)
library. 
The scaling results for Piz Daint are presented in
Figure~\ref{fig:piz-daint-scale} for four different approaches (Horovod-MPI,
gRPC, gRPC+MPI, and Horovod-MPI). It is clear that gRPC+MPI approach shows the worst
scaling. This is because of its single-threaded implementation. The
bottleneck is especially visible for NASNet-large model as a large number of
model parameters need to be transferred over the network. Baidu-MPI and
Horovod-MPI perform very similar, with Horovod-MPI showing slightly better
results. The gRPC results fall slightly behind the Baidu-MPI and Horovod-MPI
solutions. Given that all these implementations use the same underlying communication
network the performance difference can be attributed to the way the parameter
server solution is implemented. 
While training approaches have an impact on performance, the scaling behavior
is also directly related to the size of the DNN. Mobilenet shows worst scaling
compared to the ideal (16\% efficiency for Horovod-MPI), as communication of
gradients cannot be overlapped (hidden behind) the relatively smaller
computation. NASNet-large, the largest network we have tested, shows
near-ideal scaling (92\% efficiency for Horovod-MPI as the computation is large
enough to be overlapped with gradient communication. Resnet-50 falls in between
the two extremes with 71\% efficiency for Horovod-MPI. 
\MySection{Related Work}
\label{sec:related}
There are a significant number of research efforts from academia and industry
toward understanding and developing scalable deep learning systems. 
Ben-Nun and Hoefler present an in-depth concurrency analysis of various DNN
architectures with a focus on distributed training in ~\cite{distdl-preprint}.
Cui {\it et al.} propose a ``GeePS'' parameter server to support scalable
data-parallel training across GPU nodes by mitigating data movement overhead
caused by GPU stalling in~\cite{Cui:EuroSys16:GeePS}. Chilimbi {\it et al.}
propose a distributed system called Adam to achieve scalable DL training for
large DNN models, and a 120-machine cluster was used for
a demonstration~\cite{Chi:OSDI14:Adam}.
Shi {\it et al.} evaluated the performance of distributed DL frameworks such as
TensorFlow, CNTK, Caffe-MPI and MXNet over single-GPU, multi-GPU, and multi-node
(4 nodes) scenarios~\cite{dlbench,Shi:2017:modleDL}. They identify several
performance bottlenecks such as I/O access and data communication among GPUs.
However, the scalability of these DL frameworks is not evaluated.
The authors of~\cite{intel-sysml} have presented an extension of the
Horovod~\cite{horovod} design that uses the Intel Machine Learning Scaling
(MLSL) library for distributed training with TensorFlow. 
However, the study is focused on scaling CPU-based training only and no GPU
results are presented. 
\MySection{Conclusion}
\label{sec:conclusion}
In this work, we evaluated a wide range of communication methods that can be
used with the TensorFlow deep learning framework.  Our focus has been on the
scalability, and usability of the various techniques. We compared MPI based
reduction methods, custom written reduction schemes, and TensorFlow's native
parameter server model. For our evaluation, we used three different image
classification networks, each with a significantly different number of
parameters to be exchanged. %
We further introduced an improved
reduction method for GPU based data. The optimized designs have been integrated
into MVAPICH2 and have been made publicly available since the MVAPICH2-GDR 2.3rc1 release. 
This method shows up to 1.4$\times$ faster Allreduce operations compared to the NVIDIA NCCL2 communication library in
micro-benchmarks. %
Our performance evaluation shows that the customized reduction based methods, i.e., Horovod-MPI
and Horovod-NCCL, outperform the parameter server methods, i.e., gRPC and gRPC+X, 
for all three evaluated neural nets and all three systems. 
The experimental results show that the proposed optimizations help Horovod-MPI
achieve approximately 98\% and 90\% scaling efficiency for training on 16 and
64 GPU nodes on the RI2 and Owens clusters, respectively. Finally, Horovod-MPI
yields 1.8$\times$ and 3.2$\times$ higher throughput than the native gRPC method
for ResNet-50 and MobileNet, respectively, on the Piz Daint cluster across 128
nodes.
 
\vspace{1.0ex}
\section*{Acknowledgment}
This research is supported in part by National Science Foundation grants
\#CNS-1513120, \#ACI-1450440 and \#CCF-1565414, and a grant from the Swiss
National Supercomputing Centre (CSCS) under project ID s716. 
The authors would like to thank Jonathan Perkins and Dr. Khaled Hamidouche for
extending invaluable help in implementing the pointer cache design for
MVAPICH2-GDR.


\end{document}